\documentclass[twocolumn]{aastex631}
\usepackage{graphicx}
\usepackage{amsmath}
\usepackage{multirow,soul}

\shorttitle{Predicting Mercury instability using a Convolutional Neural Network}
\begin{document}

\title{AI can identify Solar System instability billions of years in advance}

\correspondingauthor{Dorian S. Abbot}
\email{abbot@uchicago.edu}

\author{Dorian S. Abbot}
\affiliation{Department of the Geophysical Sciences \\
The University of Chicago \\
Chicago, IL 60637 USA}

\author{J.D. Laurence-Chasen}
\affiliation{National Renewable Energy Laboratory \\
Golden, CO 80401 USA}
\affiliation{Research Computing Center \\
The University of Chicago \\
Chicago, IL 60637 USA}

\author{Robert J. Webber}
\affiliation{Department of Computing \& Mathematical Sciences \\
California Institute of Technology \\
Pasadena, California 91125 USA}

\author{David M. Hernandez}
\affiliation{Department of Astronomy \\
Yale University \\
New Haven, CT 06511 USA}

\author{Jonathan Weare}
\affiliation{Courant Institute of Mathematical Sciences \\ 
New York University \\ 
New York, NY 10012 USA}

%%%%%%%%%%%%%%%%%%%%%%%%%%%%%%%%%%%%%%%%%%%%%%%%%%%%%%%%%%%%%%%%%%%%
\begin{abstract}
Rare event schemes require an approximation of the probability of the rare event as a function of system state. Finding an appropriate reaction coordinate is typically the most challenging aspect of applying a rare event scheme. Here we develop an artificial intelligence (AI) based reaction coordinate that effectively predicts which of a limited number of simulations of the Solar System will go unstable using a convolutional neural network classifier. The performance of the algorithm does not degrade significantly even 3.5 billion years before the instability. We overcome the class imbalance intrinsic to rare event problems using a combination of minority class oversampling, increased minority class weighting, and pulling multiple non-overlapping training sequences from simulations. Our success suggests that AI may provide a promising avenue for developing reaction coordinates without detailed theoretical knowledge of the system. 
\end{abstract}

%%%%%%%%%%%%%%%%%%%%%%%%%%%%%%%%%%%%%%%%%%%%%%%%%%%%%%%%%%%%%%%%%%%%
\section{Motivation} \label{sec:intro}

Numerical calculations indicate that there is a $\sim$$10^{-2}$ probability that Mercury's orbit will destabilize in the next 5 billion years 
\citep{laskar1994large,laskar2009existence,zeebe2015dynamic,abbot2021rare,abbot2023simple}. \citet{abbot2021rare} applied a rare event sampling scheme \citep{webber2019practical} to the REBOUND $N$-body code \citep{rein2012rebound} and found that the probability of a Mercury instability event in the next 2 billion years is $\sim$$10^{-4}$ at a 1--2 order of magnitude reduction in computational cost. Rare event sampling schemes require a reaction coordinate that can approximate the probability of a future rare event. \citet{abbot2021rare} used physically motivated guesses pruned by a simple machine learning scheme, Lasso logistic regression \citep{tibshirani1996regression}, to select the range (maximum minus minimum) in 
Mercury-Venus Minimum Orbit Intersection Distance (MOID) over a 400~Myr period as the reaction coordinate.

Here we approach reaction coordinate development as an Artificial Intelligence (AI) supervised learning classifier problem on the Solar System simulations of \citet{abbot2023simple}. The main challenges for this task are the significant class imbalance and relative sparsity of training data. We are able to overcome these challenges using a combination of minority class oversampling, increased minority class weighting, and pulling multiple non-overlapping training sequences from simulations. We achieve similar accuracy to our previous reaction coordinate in the 2 billion years before the instability. Remarkably, the learned classification algorithm remains effective up to 3.5 billion years before the instability.

%%%%%%%%%%%%%%%%%%%%%%%%%%%%%%%%%%%%%%%%%%%%%%%%%%%%%%%%%%%%%%%%%%%%
\section{AI Model} \label{sec:model}

The model is implemented using TensorFlow v2.10.0 and our code is available at \texttt{https://knowledge.uchicago.edu/record/10055}.

We train on Mercury and Venus eccentricity time series from 46 unstable simulations and 482 stable simulations performed by \citet{abbot2023simple}, representing an order of magnitude oversampling of the minority (smaller) class, which in this case is unstable simulations. During model training we apply a weight of 10 to unstable simulations via TensorFlow's \texttt{class\_weights} parameter to further mitigate the class imbalance (much larger number of stable simulations than unstable simulations). We downsample the data so that the time spacing is 200~kyr and pull 10 sub-sequences of length 400 Myr per simulation. We use a 7-fold cross-validation scheme, where each test fold comprises sequences from 6 unstable simulations and sequences from the remaining 40 simulations make up the training fold. This means that we break the entire data set into 7 sub-sets, called folds, use one fold as the test data set and the rest of the data as the training data set, then loop through so that each fold is the test data set once. For hyperparameter tuning we use half the test fold for validation. Hyperparameters are higher-order parameters usually related to the architecture of the network or convergence scheme that must be set for a single iteration of convergence. Hyperparameter tuning involves changing these higher-order parameters and seeing if the loss function, or cost function, improves. 

We test four model architectures: LSTM (Long Short-Term Memory, a recurrent neural network), 1D CNN (convolutional neural network), 2D CNN, and XGBoost (a tree-based method). We achieve similar performance with all architectures and choose 1D CNN because it is the fastest. 
1D CNN is based on \citet{wang2017time}. It contains three 1D convolutional layers that convolve on the time dimension. Each convolutional layer is followed by a batch normalization layer and reLU activation. Batch normalization normalizes the activation inputs from the convolutional layer to zero mean and unit standard deviation, which leads to more efficient optimization of the scheme and regularization \citep[robustness of the scheme to different inputs, or reduction in overfitting to the training data, ][]{goodfellow2016deep}. ReLU stands for rectified linear unit, in which the response is linear in the signal above a threshold and zero below this threshold. This is a way to incorporate nonlinearity into the network without introducing too much complexity \citep{goodfellow2016deep}. The penultimate layer is a global average pooling layer \citep{lin2013network}, which computes the average activation of a given artificial neuron over all times. This is followed by a fully connected output layer with a softmax activation function, which converts a set of real numbers to a probability distribution weighted by the exponential of the real numbers, so that most of the probability is associated with the maximum real number \citep{goodfellow2016deep}.

We perform Bayesian optimization hyperparameter tuning on six parameters: the number of filters and the kernel size of each of the three convolutional layers. We use a maximum of 500 training epochs, or iterations of the training on all data, and stop training if loss (cost) does not improve after 50 epochs.

%%%%%%%%%%%%%%%%%%%%%%%%%%%%%%%%%%%%%%%%%%%%%%%%%%%%%%%%%%%%%%%%%%%%
\section{Results} \label{sec:results}

\begin{figure}[ht!]
\centering
\includegraphics[width=\linewidth]{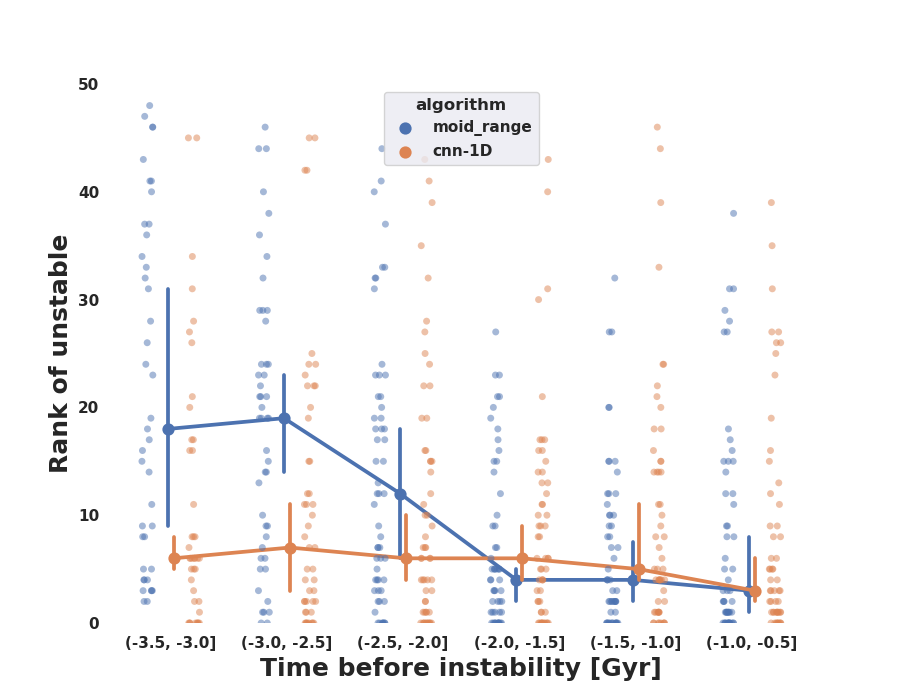}
\caption{Rank of the unstable simulation in a pool containing it and 49 stable simulations in terms of probability of instability for both the 1D CNN (orange) and MOID range (blue), which is the reaction coordinate used by \citet{abbot2021rare}. A rank of 0 is a perfect score. Small circles represent individual instances of the test and large circles are the medians of all instances in a given time range. Error bars on the large circles are 95\% confidence intervals calculated using bootstrapping. The horizontal axis shows the time range before the instability that the unstable sequence was drawn from.}
\label{fig}
\end{figure}

The \citet{webber2019practical} rare event scheme splits and kills simulations on the basis of the ranked order of the reaction coordinate. As a result, we evaluate model performance using a pool of sequences from the test data set drawn from 1 unstable and 49 stable simulations. The score is the rank of the unstable simulation in this pool, based on probability of instability, with indexing starting at 0. We compare the performance of 1D CNN and MOID range in Fig.~\ref{fig}. For up to 2 billion years before the instability both 1D CNN and MOID range reliably rank the unstable simulation among the 10 most likely to go unstable, indicating that the Solar System has a dynamical memory of variables relevant for stability three orders of magnitude larger than the Lyapunov timescale. Given that using MOID range as a reaction coordinate produced a 1-2 order of magnitude computational speed up \citep{abbot2021rare}, this indicates that our AI classification scheme is able to produce an effective reaction coordinate without physically motivated guesses. This suggests that AI classification based on a limited number of test examples is a promising avenue for development of reaction coordinates in other rare event problems. Interestingly, the behavior of 1D CNN does not significantly degrade even 3.5 billion years before the instability.  We tried to identify the physical basis for the signal the AI learned using standard explainable AI methods such as SHAP \citep{lundberg2017unified} and LIME \citep{ribeiro2016should} but results were inconclusive.

%%%%%%%%%%%%%%%%%%%%%%%%%%%%%%%%%%%%%%%%%%%%%%%%%%%%%%%%%%%%%%%%%%%%
\begin{acknowledgments}
%\vspace{1 cm}

This work was completed with resources provided by the University of Chicago Research Computing Center. This work was supported the Army Research Office, grant number W911NF-22-2-0124.
\end{acknowledgments}

\end{document}